\documentstyle[NATO,numreferences,epsfig]{crckapb}

\begin{opening}
\title{Information from the Beginning}
\subtitle{Estimating observable effects of quantum-gravity discreteness on patterns of cosmic background
radiation}  
\author{Craig J. Hogan}
\institute{Astronomy and Physics Departments \\
 University of Washington\\
 Seattle, Washington 98195, USA}
\end{opening}
 \runningtitle{}
\begin{document}
\begin{abstract}
Requiring  black hole
evaporation to be quantum-mechanically coherent
imposes a universal, finite ``holographic
bound'', conjectured to be
 due to fundamental discreteness of quantized gravity, on the amount of information 
carried by any physical system. This bound is applied to   the information budget
in the standard slow-roll model  of cosmic
inflation. A simple estimate  suggests  that when quantum gravity is included,
 fluctuations during inflation have a discrete spectrum with a limited information content, only about
$10^5$ bits per mode, fixed by the inverse scalar perturbation amplitude.
This scarcity of information may allow direct observation of  quantum-gravity eigenmodes in the
anisotropy of cosmic background radiation. 
\end{abstract}

\bigskip

{\it ``This is what we found out about Nature's book keeping system:
 the data can be written onto a surface, and the pen with which
 the data are written has a finite
size.''}

-Gerard 't Hooft

\newpage
\section{Primordial quanta}
 
Observations of the cosmic background now penetrate to a time when the
number of quanta in the universe was a relatively
 small number.  It is natural to ask whether we can learn something
from such extraordinary data about  the  fundamental quantum structure of space and time.
Here I offer  arguments that this may indeed be possible, based on an estimate of the
 information content of
the universe near the beginning.

It is now thought that cosmological perturbations   originate
as quantum fluctuations of the inflaton and the graviton fields during inflation.
Their properties are calculated using the theory of    quantum
  fields in curved spacetime  
\cite{Starobinsky:1979ty,Hawking:1982cz,Guth:1982ec,Bardeen:1983qw,Starobinsky:1982ee,Halliwell:1985eu,Grishchuk:1993ds}. 
Zero point fluctuations   in the quantum modes of the
  the inflaton field   give rise to scalar perturbations, those in the graviton  field to tensor
perturbations.

 The spatial structure of 
 field quanta in the original fluctuations ``freezes out'' as they
increase in size beyond the  inflationary event horizon, since causality prevents them
from fluctuating further.  From a quantum point of view, their states squeeze into a definite spatial
projection, and then grow coherently by an  exponential factor during the many subsequent e-foldings of
inflation, creating an enormous number of coherent quanta (or holes in the inflaton condensate), in phase with
the original quantum  wave. From the classical point of view,  the quantum fluctuations create
permanent
  perturbations in the classical gravitational gauge-invariant potential
 $\phi_m$\cite{bardeen1980},
leading to  observable background anisotropy and  large 
scale structure\cite{cobeDMR,bennett,boom,bond01}.
 
The   large-scale classical perturbations   observed today are thus a
direct result of the quantum field activity during inflation; indeed, the pattern of microwave
anisotropy on the largest scales corresponds to a faithfully amplified image  of microscopic
field configurations
 as they froze out during inflation. Roughly speaking, each
hot or cold patch on the  sky derives originally from about one quantum, in the sense that the
occupation number at freeze-out was of order unity; the energy of a fluctuating patch was typically $E\approx
hH$, where $h$ is Planck's constant and $H$ is the expansion rate (Hubble's constant) during inflation.

The model leads to
 a remarkable phenomenon:   primordial
 anisotropies that {\em are simultaneously the smallest and
largest imaged  entities in nature}.  COBE gave us our first glimpse \cite{cobeDMR,bennett} of what 
these structures look like. 
  When we look at the largest structures in the cosmic background radiation--- the
largest and  most distant things we can possibly see, stretching across the   sky at the edge of the
universe--- we are looking at images of  particle wavefunctions
   imprinted   when they were single quanta far smaller than the
 smallest subatomic structure   seen in the
lab. (The nearest laboratory analogs are images of wavefunctions of
 trapped Bose-Einstein condensates that
trace the   zero-point mode of their trap, but even these are not free vacuum fluctuations.) The universe    
acts  like a giant natural microscope, providing us with 
 a natural virtual-particle observatory.

The standard  calculation  of these processes\cite{lythriotto} uses a
semiclassical  approximation: spacetime is assumed to be classical (not quantized), and the
perturbed fields (the inflaton and graviton) are described using relativistic quantum field
theory, essentially (in the limit of  free massless fields) an infinite collection of quantized
harmonic oscillators. The Hilbert space of this system is infinite, so although the fields are
quantized, they are continuously variable functions that can assume any values. 
The theory
generically predicts random-phase gaussian noise with a continuous spectrum determined by the parameters of the
inflaton potential.  
 Although sky maps contain images of ``single quanta,''  the amount of information they carry is in principle
infinite.

It has always been acknowledged that this description is incomplete, and will be
modified by including a proper account of spacetime quantization.  
  Although the fundamental theory of quantum gravity is not known, a ``holographic entropy bound'' already 
constrains with remarkable precision the total number of fundamental quantum degrees of freedom.
  The complete Hilbert space of a bounded volume is
  finite and discrete  rather than infinite and continuous, limiting
the range of accessible configurations in any region to a definite, calculable number.   In particular
 this limit
applies to  the coupled system of inflaton-field and spacetime configurations, together with all their
possible fluctuations, during inflation. Like bound states of an atom, 
 but unlike a free quantum field or harmonic oscillator, the spectrum of states is discrete; indeed the
number of  eigenstates, unlike   atoms, is even finite.

   A simple estimate\cite{hogan02} suggests
that  in standard inflation, 
  the amount of information in the anisotropy is remarkably limited: it can be
  described with only about $10^5$
bits per sky-harmonic mode, implying that the  perturbations
should be ``pixelated'' in some way. In principle, this effect may be observable,
 and provide concrete data on the discrete elements or
eigenstates of 
 quantum gravity.

\section{The Holographic Principle}

Hawking radiation from black holes presents a conundrum in information accounting.
Matter falling into the hole travels to the central singularity long before its energy is
radiated. It appears that either the information  of infalling matter is lost, in contradiction with
quantum mechanics, or else the black hole itself has coherent quantum states that store the information
between the time of infall and the time of particle evaporation.

The latter point of view is the one consistent with quantum mechanics and seems to be correct, but it
has radical consequences.\cite{thooft93,susskind95,thooft00,bousso02} Hawking's calculation of black hole
evaporation showed that vacuum states of particles incoming from  the distant past of a black hole spacetime
become populated with a thermal spectrum in the far future, with a precisely calculable entropy. If quantum
mechanics holds, then all the processes of black hole formation and evaporation should be time reversible.
 The evaporation products of a black hole can be time and parity reversed, then run
backwards in time, to form a black hole that will slowly grow and eventually fly apart by throwing out
whatever macroscopic objects (TV sets, astronauts, whatever)  were thrown in originally. This 
time-reversed situation is also physical, it is just extremely unlikely. The example shows how the
laws of gravity can be  essentially statistical in nature, even for a black hole.

Assuming this coherent behavior, the evaporating
 black hole system allows a complete accounting of the
number of internal  quantum states of the black hole: it must be sufficient to store all the information
needed for   the radiated particles. 
 A black hole, including its quantum-gravitational and   particle degrees of
freedom, has a density of states given by
$e^S$, where    the entropy $S$  depends only on the area $A$ of the event horizon
 of the hole: $S=A/4$ in Planck units.
Roughly speaking, each radiated particle carries about a bit of
information; when it is radiated, it reduces the mass of the hole by about $M^{-1}$ (with a wavelength at
infinity of order the Schwarzschild radius, $M$), and reduces the area of the event horizon (and the entropy)
by about one Planck area.
Since a black hole is the maximal entropy state for any spatial region of a given bounding-area size,
this result generalizes to a bound on the information content of any physical system.

Therefore, the Hilbert space of everything
 is finite and discrete: for everything in  a three-dimensional region\footnote{The three-dimensional regions
in the most general formulation are null sheets.} bounded  by
a two-dimensional
 surface of area $A$, the complete
quantum state is specified by at most
 $n=A/4 \ln 2$ binary numbers, and there are at most $2^n$ distinguishable outcomes of any experiment.
The  
 information content is the same as   
$n$ binary spins, corresponding to one spin per $0.724 \times 10^{-65}\ {\rm cm}^2$ of bounding area.
(Since the spins are
in general entangled, this is most correctly described as
$n$ quantum bits, or ``qubits'', of information). 
Field theory, even a single harmonic oscillator, has an infinite Hilbert space, so the holographic
entropy bound imposes radical constraints on physics.
Discreteness and nonlocality apparently exist in nature  that
are not modeled in field theory.

\section{ Holography and Cosmology}

If the holographic bound is correct, the observable universe today has at most
$ 3\pi/\Lambda \ln 2\approx 10^{120}$  qubits of information, where $\Lambda$ is the current value of
the cosmological constant. Although this is a lot less than infinity, nobody has yet suggested a way
that such a large number of degrees of freedom could be distinguished in practice from a continuous
system (see e.g. \cite{lloyd}). 

On the other hand, the maximum information within the event horizon during inflation was much smaller: 
\begin{equation}
S_{max}=\pi/H^2. 
\end{equation}
We know from the cosmic  background maps that $H$ was less than about $10^{-5}$ 
(because graviton fluctuations produce tensor perturbations today with $\delta T/T\approx H$), but even at this
level the total information content is only $10^{10}$ qubits; all possible physical situations could be
encoded on a laptop computer.  

A more detailed  calculation,\cite{hogan02} based on the specific context of ``slow-roll inflation'', 
suggests that for observable fluctuations,
the amount of information 
 is even less. 
The basic assumption of this calculation is a ``discreteness ansatz'' that the  inflationary universe system---
inflaton plus gravity--- makes discrete transitions between states separated by one bit of entropy. This would
follow naturally
 if   the  reason for the existence of the holographic bound is that the fundamental theory
contains discrete elements. It  seems plausible that this is the case, given recent 
development of candidates for fundamental theory such as M theory and loop quantum gravity\cite{ashtekar02}
that display discrete and holographic features. It also seems natural given the nature of particle-emission
processes in the black hole system, where  emission of each particle is associated with reduction by about
one bit in $S$.

The classical evolution of slow-roll inflation then corresponds, from the quantum point of view, to adding
binary qubits one at a time. 
The  key assumption   is that  {\it the background spacetime  and   horizon-scale
  fluctuations make
   transitions between discrete   states, each of which adds (or subtracts)
  one bit of information  to the total maximum observable entropy.}
That is,   $H$   comes in discrete
steps of $n$:
\begin{equation} \label{eqn:minimal}
H_i=\sqrt{\pi\over n_i \ln 2}
\end{equation}
where $n_i$ are integers. 
The values of $H$ and the inflaton $\phi$ on the horizon scale are connected by the usual classical equations
of inflation, so $\phi$ also takes on  discrete values.

The     classical expectation value $\phi_c$ of the inflaton
  field and the inflationary expansion rate $H$ obey
 the
Friedmann equation controlling the expansion,
\begin{equation} \label{eqn:friedmann}
H^2={8\pi \over 3}\left[V(\phi_c)+ {\dot\phi_c^2/2}\right].
\end{equation}
and the  dynamical equation for $\phi_c$,
\begin{equation}
\ddot\phi_c+3H\dot\phi_c+V'(\phi_c)=0,
\end{equation}
where the
effective potential is
$V(\phi_c)$ and $V'\equiv dV/d\phi_c$. \cite{lythriotto}
For definiteness, assume that the observed modes crossed the
inflationary horizon  during a standard so-called ``slow roll''  or Hubble-viscosity-limited phase
of inflation, corresponding to $V'/V\le\sqrt{48\pi}$ and $V''/V\le 24\pi$, during which 
  $\dot\phi_c\approx -V'/3H$.
The rate of the roll is much slower than the expansion rate $H$, so the kinetic term in 
 Eq.  (\ref{eqn:friedmann}) can be ignored in   the mean evolution.
The 
  slow-roll phase of inflation creates    approximately
scale-invariant curvature perturbations. 

We define a
 combination of inflationary parameters   by
\begin{equation}
Q_S\equiv {H^3\over|V'|} \approx \left({V^3\over V'^2}\right)^{1/2} \left({8\pi\over 3}\right)^{3/2}.
\end{equation}
This combination   can be estimated
fairly accurately, since it also controls the amplitude of the   scalar perturbations observed in the microwave
background anisotropy. The best fit to the
four-year COBE/DMR data, assuming scale invariance ($n=1$)  and zero amplitude tensor modes, yields
\cite{bennett,hogan02}
\begin{equation}
Q_S=9.4\pm 0.84\times 10^{-5}.
\end{equation} 

Putting this together one finds that there is a steady increase in observable entropy at a rate
\begin{equation} \label{eqn:rate}
\dot S_{max}= {8\pi^2\over 9}{V'^2\over H^5}
=8\pi^2 H Q_S^{-2}.
\end{equation}
Every inflationary $e$-folding, $S_{max}$ increases by an amount of order $10^{10}$ due  to the
classical evolution of the system as $H$ slowly decreases.

However, the information attached to the observable quantity--- the
horizon-scale perturbation in  the inflaton--- is much smaller than this. 
The  variation in total entropy
associated with a horizon-scale perturbation
$\delta\phi $ is:
\begin{equation}
\delta S_{tot}={8\pi^2\over 3} \left[{\delta\phi\over H}\right] Q_S^{-1}.
\end{equation}
The   standard field theory
 analysis   for the horizon-size
perturbations tells us that 
the quantity $[\delta \phi/H]$ is   statistically determined, with a   continuous
gaussian statistical distribution of order unit width.
The corresponding increment $\delta S \approx   (8\pi^2/3) Q_S^{-1} \approx 10^5$ then is roughly the jump in
the total observable cosmological entropy associated with the  
horizon-scale inflaton perturbations.  
This is much less than the increase given by Eq. (\ref{eqn:rate}) in the total information 
during an expansion time, of order $Q_S^{-1}$ rather than  $Q_S^{-2}$. Presumably this can be attributed to
the fact that almost all of the entropy growth is associated with the classical slow roll of the inflaton
condensate, which leaves no imprint in the anisotropy.

Thus for any value of the inflationary Hubble constant, the 
horizon-size inflaton perturbations, which are the degrees of freedom frozen into our sky, contain only 
about
$10^5$ bits per mode. For example, the fluctuations seen in the COBE map, with about 1000 pixels,
intrinsically contain only about
$10^8$ bits or 10 Megabytes.
This  could in principle  be encoded in a color laptop display screen with no loss of
information. If we had a dataset with enough resolution and a high enough signal-to-noise ratio, and
knew the ``encoding'' (the projection of the quantum-gravity eigenstates onto the sky), we would find
that the data could be fit surprisingly well using these modes, and that the fit would not improve with
the addition of additional parameters. If there is substantial redundancy in the encoding, the number of
distinguishable eigenstates could  be much less than this limit, so a statistical search might find
quantum-gravity signatures even in current datasets.

To put the same point more poetically: when the  letters   of the writing on the sky are known,  the pattern
will no longer appear as  a meaningless jumble of random noise, and the significance of the whole pattern will
be interpreted completely and transparently in terms of  those letters--- the eigenmodes of the inflationary
system in  fundamental theory. When that is done, we will know everything it is possible to know about the
beginning.  All we have done here is estimate how many letters there are.

\acknowledgements
This work was supported by NSF grant AST-0098557 at the University of Washington.

{}

\end{document}